# Message and State Cooperation in a Relay Channel When the Relay Has Strictly Causal State Information

Min Li[1], Osvaldo Simeone[2] and Aylin Yener[1]

[1]Dept. of Electrical Engineering, The Pennsylvania State University, University Park, PA 16802
[2]Dept. of Electrical and Computer Engineering, New Jersey Institute of Technology, University Heights, NJ 07102
mxl971@psu.edu, osvaldo.simeone@njit.edu, yener@ee.psu.edu

*Abstract*—A state-dependent relay channel is studied in which strictly causal channel state information is available at the relay and no state information is available at the source and destination. Source and relay are connected via two unidirectional out-of-band orthogonal links of finite capacity, and a state-dependent memoryless channel connects source and relay, on one side, and the destination, on the other. Via the orthogonal links, the source can convey information about the message to be delivered to the destination to the relay while the relay can forward state information to the source. This exchange enables cooperation between source and relay on both transmission of message and state information to the destination. First, an achievable scheme, inspired by noisy network coding, is proposed that exploits both message and state cooperation. Next, based on the given achievable rate and appropriate upper bounds, capacity results are identified for some special cases. Finally, a Gaussian model is studied, along with corresponding numerical results that illuminate the relative merits of state and message cooperation.

## I. INTRODUCTION

In a wireless network, the main impediments to reliable communications are usually fading and interference. To analyze the performance limits of channels in the presence of fading and interference, a conventional model assumes that the channel is affected at each time instant by a state variable, which is controlled by a certain state distribution, and accounts for fading and/or interference [1]–[3]. State-dependent channels are usually classified on the basis of the availability of channel state information at encoders and decoders. Specifically, transmitting nodes may have no state information, or else be informed about the state sequence in a strictly causal, causal, or non-causal way [3]–[5]. For decoders, it is enough to distinguish between the case of state information or no state information available [3].

In previous work, capacity-achieving strategies have been proposed for point-to-point memoryless channels with non-causal [2], or causal [1] state information at the encoder and no state information at the decoder. These results, and the ones discussed throughout the paper, assume that the state sequence is independently and identically distributed (i.i.d.).

The work of M. Li and A. Yener was supported in part by the National Science Foundation under Grants CNS 0721445, CNS 0964364, CCF 0964362 and DARPA ITMANET Program under Grant W911NF-07-1-0028. The work of O. Simeone was supported in part by the National Science Foundation under Grant CCF 0914899.

Several multi-user channels have also been widely investigated in similar settings and a non-exhaustive list includes multiple access channels (MACs) [4]–[8] and relay channels [9], [10]. For the relay channel, reference [10] investigates the case of non-causal state information at the relay, and proposes a coding scheme that combines the strategies of decode-and-forward [11] and precoding against the state, while reference [9] studies the case of causal state information at the relay, and derives achievable rates by combining the ideas of compress-and-forward [11] and adapting input codewords to the state (also known as Shannon strategies [1]).

This work also focuses on a state-dependent relay channel, but unlike [9], [10], assumes that state information is available at the relay in a *strictly causal* fashion, i.e., the state sequence at a given time is known up to the previous instant at the relay. This scenario is more relevant in practical scenarios. For instance, an interfering sequence, caused by other users' transmission, can be learned as it is observed, and thus in a strictly causal manner. With strictly causal state information, the strategies leveraged in [9], [10] of precoding against the state or Shannon strategies cannot be applied. More fundamentally, the question arises as to whether strictly causal, and thus outdated, state information may be useful at all in a memoryless channel with i.i.d. state sequence. In fact, it is well known that strictly causal state information is useless in point-to-point channels [12].

Recently, in [4], [5], it was found that for two-user MACs with independent or common state information available strictly causally at the encoders, unlike for point-to-point channels, capacity gains can be accrued by leveraging strictly causal state information at the encoders. Our recent work [13] further extended such results to MACs with arbitrary number of users by proposing a coding scheme inspired by noisy network coding [14]. In [4], [5], [13], the main idea is to let each transmitter convey a compressed version of the outdated state information to the decoder, which in turn exploits such information to perform partially coherent decoding. The results show that an increase in the capacity region can be obtained by devoting part of the transmission resources to the transmission of the compressed state.

In this work, we focus on state-dependent relay channels and study the performance trade-off arising from the need to send both message and state information from source and relay to destination. Specifically, we consider a three-node relay channel where the source and relay are connected via two out-of-band orthogonal links of finite capacity, and a state-dependent memoryless channel connects the source and relay, on one side, and the destination, on the other. Source and destination have no state information, while the relay has access to the state information in a strictly causal manner. The channel model is depicted in Fig. 1. This model is related to the class of relay channels, that are not state-dependent, with orthogonal links from the source to the relay and from the source and relay to the destination investigated by El Gamal and Zahedi [15]. In fact, in the scenario under study, we simplify the source-to-relay link by modeling as a noiseless finite-capacity link, while adding a similar relay-to-source link. Cooperation as enabled by orthogonal noiseless links, also referred to as conferencing, was first introduced by Willems [16] for a two-user MAC channel. It is noted that, in practice, orthogonal links can be realized if nodes are connected via several different radio interfaces or wired links [17].

In the considered model, cooperation between source and relay through the conferencing links can aim at two distinct goals: $i$) Message transmission: Through the source-to-relay link, the source can provide the relay with some information about the message to be conveyed to the destination, thus enabling message cooperation; $ii$) State transmission: Through the relay-to-source link, the relay can provide the source with some information about the state, thus enabling cooperative transmission of the state information to the destination. We propose a transmission scheme inspired by noisy network coding and establish the corresponding achievable rate. Moreover, based on the given achievable rate, we identify capacity results for some special cases of the considered model. Finally, we present achievable rates and some capacity results for the Gaussian version of the system at hand and elaborate on numerical results. Due to space limitation, most of the proofs are omitted and can be found in [18].

*Notation*: Probability distributions are identified by their arguments, e.g., $p_X(x) = \Pr[X = x] \triangleq p(x)$. $x^i$ denotes vector $[x_1, ..., x_i]$. $\mathbb{E}[X]$ denotes the expectation of random variable $X$. $\mathcal{N}(0, \sigma^2)$ denotes a zero-mean Gaussian distribution with variance $\sigma^2$. $\mathcal{C}(x)$ is defined as $\mathcal{C}(x) = \frac{1}{2} \log_2(1 + x)$.

## II. SYSTEM MODEL

In this section, we formalize our relay channel model and give relevant definitions. As depicted in Fig. 1, we study a three-node relay channel where the source and relay are connected via two unidirectional out-of-band orthogonal links of finite capacity, while there is a state-dependent memoryless channel between the source and relay, on one side, and the destination, on the other. Note that the relay transmits and receives simultaneously over two orthogonal channels.

The channel is characterized by the tuple:

$$(\mathcal{X} \times \mathcal{X}_R, \mathcal{S}, \mathcal{Y}, p(s), p(y|s, x, x_R), C_{SR}, C_{RS}) \quad (1)$$

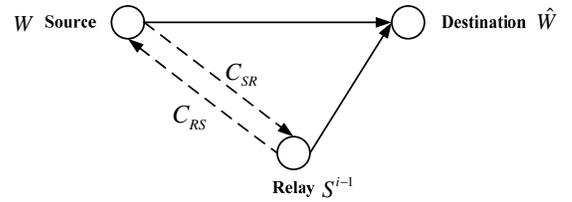

Fig. 1. A state-dependent relay channel with two unidirectional out-of-band orthogonal links.

with source input alphabet $\mathcal{X}$, relay input alphabet $\mathcal{X}_R$, destination output alphabet $\mathcal{Y}$ and channel state alphabet $\mathcal{S}$. The capacity per channel use of the source-to-relay and relay-to-source out-of-band, also known as conferencing [16], links are given by $C_{SR}$, $C_{RS}$ respectively. The state sequence is assumed to be i.i.d., i.e., $p(s^n) = \prod_{i=1}^{n} p(s_i)$. The relay channel is discrete memoryless (DM) in the sense that at any discrete time $i = 1, ..., n$, we have

$$p(y_i | s^i, x^i, x_R^i, y^{i-1}) = p(y_i | s_i, x_i, x_{R,i}). \quad (2)$$

We assume that the state information is available to the relay in a *strictly causal* manner while there is no state information at the source and destination.

*Definition 1:* Let $W$, uniformly distributed over the set $\mathcal{W} = [1 : 2^{nR}]$, be the message sent by the source. A $(2^{nR}, n)$ code consists of:

1) Conferencing codes: Conferencing mappings are defined as

$$h_{SR,i} : \mathcal{W} \times \mathcal{T}_{RS}^{i-1} \to \mathcal{T}_{SR,i}, \quad (3)$$
$$h_{RS,i} : \mathcal{S}^{i-1} \times \mathcal{T}_{SR}^{i-1} \to \mathcal{T}_{RS,i}, \quad (4)$$

where (3) generates the $i$th symbol sent on the source-to-relay link based on the message and all symbols previously received from the relay, while (4) generates the $i$th symbol sent on the relay-to-source link based on the states up current time and all symbols previously received from the source. Note that at each time $i$, $\mathcal{T}_{SR,i}$ and $\mathcal{T}_{RS,i}$ are the alphabets of the conferencing message sent from the source to relay and from the relay to source, respectively. Such mappings are permissible if the following capacity-conserving conditions are satisfied:

$$\frac{1}{n} \sum_{i=1}^{n} \log_2 |\mathcal{T}_{SR,i}| \leq C_{SR}, \frac{1}{n} \sum_{i=1}^{n} \log_2 |\mathcal{T}_{RS,i}| \leq C_{RS}. \quad (5)$$

2) Encoder mappings at the source:

$$f_i : \mathcal{W} \times \mathcal{T}_{RS}^{i} \to \mathcal{X}_i, \forall\ i = 1, ..., n, \quad (6)$$

which generates the channel input at the source at time $i$ based on the message and the information received from the relay up and including time $i$.

3) Encoder mappings at the relay:

$$f_{R,i} : \mathcal{S}^{i-1} \times \mathcal{T}_{SR}^{i} \to \mathcal{X}_{R,i}, \forall\ i = 1, ..., n, \quad (7)$$

which generates the channel input at the relay at time $i$ based on the strictly causal state information and the information received from the source up and including time $i$.

4) Decoder mapping at the destination:
$$g : \mathcal{Y}^n \to \mathcal{W}, \qquad (8)$$

which produces the estimate of message at the destination based on the received sequences.

The average probability of error, $\Pr(E)$, is defined by:
$$\Pr(E) = \frac{1}{2^{nR}} \sum_{w=1}^{2^{nR}} \Pr\left(g\left(y^n\right) \neq w \,|\, w \; sent\,\right). \qquad (9)$$

A rate $R$ is achievable if there exists a sequence of codes $(2^{nR}, n)$ as defined above such that the probability of error $\Pr(E) \to 0$ as $n \to \infty$. The capacity of this channel is the supremum of the set of all achievable rates.

## III. ACHIEVABLE SCHEME AND UPPER BOUND

In this section, we demonstrate a transmission scheme that exploits both message and state cooperation between source and relay. We also identify an upper bound on the capacity.

### A. Achievable Scheme: Burst Message Cooperation and Block-based State Cooperation

*Proposition 1:* For the DM state-dependent relay channel of Fig. 1, any non-negative rate smaller than $R$ is achievable where
$$R = \max_{\mathcal{P}} \min \begin{pmatrix} I\left(X; Y \,|\, X_R, V, U\right) + C_{SR}, \\ I\left(X, X_R, V; Y\right) - I\left(V; S \,|\, X_R, U\right), \\ I\left(X, X_R, V; Y \,|\, U\right) + C_{SR} \\ \quad + C_{RS} - I\left(V; S \,|\, X_R, U\right) \end{pmatrix} \qquad (10)$$

with the maximum taken over the distributions in the set of
$$\mathcal{P} = \big\{ p\left(v, u, s, x, x_R, y\right) : \\ p\left(s\right) p\left(v \,|\, s, x_R, u\right) p\left(u\right) p\left(x \,|\, u\right) p\left(x_R \,|\, u\right) p\left(y \,|\, s, x, x_R\right) \big\}. \qquad (11)$$

*Sketch of Proof:* Inspired by noisy network coding in [14], the same message $w$, $w \in \left[1 : 2^{nbR}\right]$, is sent at the source over all $b$ blocks of transmission with each consisting of $n$ channel uses. Thus, message information exchange between source and relay takes place only one at the beginning of the first block. This way, the source shares part of the message $w$ with the relay in order to enable message cooperation. As for the state, at the end of each block, the relay compresses the state sequence over the block *without explicit Wyner-Ziv coding*, that is, without binning as in [14]. Exchange of state information between relay and source takes place before the beginning of each block. Source and relay cooperatively send the message and state information they share, while the source sends the remaining part of the message independently and the relay sends the remaining part of the compression index alone for each block. This transmission scheme is referred to as burst message cooperation and block-based state cooperation strategy. At the end of $b$ blocks of transmission, the destination performs *joint decoding over all blocks* of reception *without explicitly decoding the compressed state information* as for the noisy network coding scheme [14]. ∎

*Remark 1:* To interpret (10) to (11) in light of the transmission strategy discussed above, we remark that $V$ represents the compressed state information and $U$ accounts for the codeword transmitted cooperatively by the source and relay, which conveys both state and message information they share. The mutual information terms in (10), in particular the conditioning on $V$, account for the fact that the destination has information about the channel via the compressed state $V$, which allows for partial or complete coherent decoding. Moreover, the second and third term in (10) reflect the cost in terms of rate to be paid for the transmission of compressed state information. □

### B. An Upper Bound

Now we present a simple upper bound.

*Proposition 2:* For the DM state-dependent relay channel of Fig. 1, the capacity is upper bounded by
$$R_{upp} = \max_{\mathcal{P}_{upp}} \min\left(I\left(X, X_R; Y\right), I\left(X; Y \,|\, X_R, S\right) + C_{SR}\right) \qquad (12)$$

with the maximum taken over the distributions in the set of
$$\mathcal{P}_{upp} = \left\{p\left(s, x, x_R, y\right) : p\left(s\right) p\left(x, x_R\right) p\left(y \,|\, s, x, x_R\right)\right\}. \qquad (13)$$

*Remark 2:* The upper bound (12) is essentially a cut-set bound [12], where the first term corresponds to the MAC cut between source-relay and destination, and the second term is the cut between source and relay-destination. □

## IV. SPECIAL CASES AND CAPACITY RESULTS

In this section, we consider three special cases of the general model studied above, namely: $i$) No message and state cooperation, in which $C_{SR} = C_{RS} = 0$; $ii$) Message cooperation only, in which $C_{SR} > 0, C_{RS} = 0$; $iii$) State cooperation only, in which $C_{SR} = 0, C_{RS} > 0$. We establish capacity results for a special class of channels for each case.

### A. No Message and State Cooperation

With $C_{SR} = C_{RS} = 0$, a general achievable rate can be identified through $R$ of (10) by setting $U = \emptyset$, since no information is shared between the source and relay. This rate turns out to be optimal, i.e., capacity-achieving, for a special class of relay channels, which includes modulo-additive state-dependent relay channels, see Example 1.

*Proposition 3:* Let $\mathcal{P}_1^*$ denote the set of distributions defined by:
$$\mathcal{P}_1^* = \left\{p\left(s, x, x_R, y\right) : p\left(s\right) p\left(x\right) p\left(x_R\right) p\left(y \,|\, s, x, x_R\right)\right\}. \qquad (14)$$

If $C_{SR} = C_{RS} = 0$,
$$H\left(Y \,|\, X, X_R, S\right) = 0, \qquad (15)$$
$$\text{and } H\left(S \,|\, X, X_R, Y\right) = 0 \qquad (16)$$

are satisfied for all distributions in $\mathcal{P}_1^*$, then the capacity is given by:
$$C_1 = \max_{\mathcal{P}_1^*} \min\left(H\left(Y \,|\, X_R, S\right), I\left(X, X_R; Y\right)\right). \qquad (17)$$

*Remark 3:* Condition (15) basically states that, when fixed $X$ and $X_R$, there is no other source of uncertainty in the observation $Y$ beside the state $S$. Condition (16), instead, says that the state $S$ is perfectly determined when $Y, X, X_R$ are known. These conditions guarantee that providing information about the state directly reduces the uncertainty about the inputs $X$ and $X_R$. The fact that the relay can increase the achievable rate up to $I(X, X_R; Y)$ in (17) can be interpreted in light of this fact since the relay signal $X_R$ directly contributes to the achievable rate even though the relay is not aware of the message by the source. □

*Example 1:* Consider a binary modulo-additive state-dependent relay channel defined by $Y = X \oplus X_R \oplus S$, where $S \sim Bernoulli(p_s)$. Let us further impose the cost constraints on the source and relay codewords $(x^n, x_R^n)$, $\frac{1}{n}\sum_{i=1}^{n}\mathbb{E}[X_i] \leq p, \frac{1}{n}\sum_{i=1}^{n}\mathbb{E}[X_{R,i}] \leq p_r$ with $0 \leq p, p_r \leq \frac{1}{2}$. Extending the capacity result of Proposition 3 to channels with cost constraints is straightforward and leads simply to limiting the set of feasible distributions (14) by imposing the constraints that $\mathbb{E}[X] \leq p$ and $\mathbb{E}[X_R] \leq p_r$, see, e.g., [19, Lecture 3]. Therefore the capacity is given by:

$$C_{\text{bin}} = \min\left(H_b(p), H_b(p * p_r * p_s) - H_b(p_s)\right), \quad (18)$$

where $p_1 * p_2 = p_1(1-p_2) + p_2(1-p_1)$, and $H_b(p) = -p\log_2 p - (1-p)\log_2(1-p)$.

As a specific numerical example, setting $p = p_r = 0.15$ and $p_s = 0.1$, we have $C_{\text{bin}} = 0.4171$. Note that without state information at the relay, the channel can be considered as a relay channel with reversely degraded components in [11]. In this case, the best rate achieved is given by [11, Theorem 2]:

$$C_{\text{bin, no SI}} = \max_{p(x)} \max_{x_R} I(X; Y | X_R = x_R) \quad (19)$$
$$= H_b(p * p_s) - H_b(p_s) \quad (20)$$
$$= 0.2912. \quad (21)$$

Hence $C_{\text{bin}} > C_{\text{bin, no SI}}$, which assesses the benefit of state information known at the relay even in a *strictly causal* manner.

*Remark 4:* The channel discussed in Example 1, has a close relationship with the modulo-additive state-dependent relay model considered by Aleksic, Razaghi and Yu in [20]. Therein, the relay observes a corrupted version of the noise (state) *non-causally* and has a *separate and rate-limited digital link* to communicate to the destination. For this class of channels, a compress-and-forward strategy is devised and shown to achieve capacity. Unlike [20], the relay obtains the state information noiselessly, strictly causally and the relay-to-destination link is non-orthogonal to the source-to-destination link. We have shown in Proposition 3 that in this case, the proposed scheme achieves capacity. □

### B. Message Cooperation Only

With $C_{RS} = 0$, the model at hand is similar to the one studied by El Gamal and Zahedi [15], where capacity was obtained for a *state-independent* channel in which a general noisy channel models the source-to-relay link. By setting $S = V = \emptyset$ and $C_{RS} = 0$ in (10), we recover a special case of the capacity obtained in [15] with noiseless source-to-relay link.

For *state-dependent* channels, a general achievable rate can be obtained through $R$ in (10) by setting $C_{RS} = 0$. Moreover, when the source-to-relay conferencing capacity $C_{SR}$ is large enough, we are able to characterize the capacity as follows. Notice that this capacity result holds for an arbitrary $C_{RS}$, not necessary $C_{RS} = 0$.

*Proposition 4:* Let $\mathcal{P}_2^*$ denote the set of distributions defined by:

$$\mathcal{P}_2^* = \{p(s, x, x_R, y) : p(s) p(x, x_R) p(y|s, x, x_R)\}. \quad (22)$$

If $C_{SR} \geq \max_{\mathcal{P}_2^*} I(X, X_R; Y)$ and arbitrary $C_{RS}$, the capacity $C_2$ is given by:

$$C_2 = \max_{\mathcal{P}_2^*} I(X, X_R; Y), \quad (23)$$

and is achieved by message cooperation only.

*Remark 5:* The capacity identified above is the same as without any state information at the relay. This result implies that when the relay is cognizant of the entire message, message transmission always outperforms sending information about the channel states. This can be seen as a consequence of the fact that in a point-to-point channel, no gain is possible by exploiting availability of strictly causal state information. □

### C. State Cooperation Only

If $C_{SR} = 0$, no cooperative message transmission is allowed. However, through the conferencing link of capacity $C_{RS}$, cooperative state transmission between the relay and source is still feasible. A general achievable rate can be identified from $R$ in (10) by setting $C_{SR} = 0$. Specifically, when $C_{RS}$ is large enough, we have the following corollary.

*Corollary 1:* Let $\mathcal{P}_3$ denote the set of distributions defined by:

$$\mathcal{P}_3 = \{p(s, v, x, x_R, y) : \\ p(s) p(v|s, x_R) p(x, x_R) p(y|s, x, x_R)\}. \quad (24)$$

If $C_{SR} = 0$ and $C_{RS} \geq \max_{\mathcal{P}_3} I(X_R; Y)$, any non-negative rate smaller than $R_3$ is achievable where

$$R_3 = \max_{\mathcal{P}_3} \min \begin{pmatrix} I(X;Y|X_R, V), \\ I(X, X_R, V; Y) - I(V; S|X_R) \end{pmatrix}. \quad (25)$$

This rate gives the capacity for the special class of relay channels characterized by (15)–(16).

*Proposition 5:* Let $\mathcal{P}_3^* = \mathcal{P}_3$ as defined by (24). If $C_{SR} = 0$, $C_{RS} \geq \max_{\mathcal{P}_3^*} I(X_R; Y)$ and (15)–(16) are satisfied for all distributions in $\mathcal{P}_3^*$, then the capacity is given by:

$$C_3 = \max_{\mathcal{P}_3^*} \min\left(H(Y|X_R, S), I(X, X_R; Y)\right). \quad (26)$$

*Remark 6:* Compared to the capacity result provided in Proposition 3 for the same class of channels (15)–(16), $C_3$ is potentially larger because a general input distribution is admissible instead of the product input distribution due to state cooperation. The resulting cooperative gain will be further discussed for the Gaussian model in Section V. □

## V. GAUSSIAN MODEL

In this section, we briefly study the Gaussian model depicted in Fig. 1, in which the destination output $Y_i$ at time instant $i$ is related to the channel input $X_i$ from the source, $X_{R,i}$ from the relay, and the channel state $S_i$ as

$$Y_i = X_i + X_{R,i} + S_i + Z_i, \quad (27)$$

where $S_i \sim \mathcal{N}(0, P_S)$ and $Z_i \sim \mathcal{N}(0, N_0)$, are i.i.d., mutually independent sequences. The channel inputs from the source and relay satisfy the following average power constraints

$$\frac{1}{n}\sum_{i=1}^n \mathbb{E}\left[X_i^2\right] \le P, \quad \frac{1}{n}\sum_{i=1}^n \mathbb{E}\left[X_{R,i}^2\right] \le P_R. \quad (28)$$

For this Gaussian model, a general achievable rate $R^G$ can be obtained from rate (10) by properly choosing Gaussian input signals such that (28) is satisfied and generating $V$ as $V = S + Q$ with $Q \sim \mathcal{N}(0, P_Q)$ for some compression variance $P_Q \ge 0$.

*Remark 7:* If the relay ignores the available state information, it only cooperates with the source in sending the message information. An achievable rate corresponding to this situation is given by

$$R^G_{\text{no SI}} = \max_{0 \le \alpha \le 1} \min \begin{pmatrix} \mathcal{C}\left(\frac{(1-\alpha)P}{N_0+P_S}\right) + C_{SR}, \\ \mathcal{C}\left(\frac{P+P_R+2\sqrt{\alpha P P_R}}{N_0+P_S}\right) \end{pmatrix}. \quad (29)$$

This rate will be later used for performance comparison. □

### A. Special Cases and Capacity Results

Now we focus on the special case where $N_0 = 0$. We first consider the case with no both message and state cooperation, following Proposition 3.

*Corollary 2:* If $N_0 = 0$ and the conferencing links satisfy $C_{SR} = C_{RS} = 0$, the capacity is given by:

$$C^G_{\text{no coop}} = \mathcal{C}\left(\frac{P+P_R}{P_S}\right). \quad (30)$$

*Remark 8:* The capacity result indicates that strictly causal state information at the relay can provide power gain for the channel considered, even though the relay knows nothing about the message information intended for destination from the source. In fact, when $N_0 = 0$, state conveying from the relay to destination can be considered as equivalently sending partial message for the source, as discussed in Remark 3. □

Next, we consider the optimality of state and message cooperation only following Proposition 4 and 5.

*Corollary 3:* If $N_0 = 0$ and the conferencing links satisfy $C_{RS} \ge \mathcal{C}\left(\frac{P+P_R+2\sqrt{PP_R}}{P_S}\right)$ with arbitrary $C_{SR}$, the capacity is given by:

$$C^G = \mathcal{C}\left(\frac{P+P_R+2\sqrt{PP_R}}{P_S}\right), \quad (31)$$

and is achieved by state cooperation only. Moreover, if $N_0 = 0$ and the conferencing links satisfy $C_{SR} \ge C^G$ with arbitrary $C_{RS}$, the capacity is also given by (31), and is attained by message cooperation only.

*Remark 9:* Example 1 in [4] implies that, if the source knows the state information as well, then the maximum rate is given by (31). Corollary 3 then quantifies the minimum capacity $C_{RS}$ necessary for this result to be attained on the relay channel of Fig. 1 where the source is not given the state information. □

From Corollary 3, we immediately have the following.

*Corollary 4:* If $N_0 = 0$, and both $C_{RS}$ and $C_{SR}$ are large enough, both state and message cooperation only are optimal and achieve the full cooperation bound (31). Compared to the case without any cooperation of (30), they both provide cooperative gain.

### B. Numerical Results and Discussions

We now provide some numerical results. We start from the special case with $N_0 = 0$ studied in Corollary 3. We first consider the performance for message cooperation only, i.e., $C_{RS} = 0$. In Fig. 2 (a), we plot the achievable rates versus conferencing capacity $C_{SR}$. We also plot the rate $R^G_{\text{no SI}}$ in (29) that is achieved when the relay does not use the available side information. It can be seen that, if $C_{SR}$ is large enough, the proposed scheme achieves the upper bound (31) and the optimal strategy is to let the relay ignore the state information as provided in Corollary 3. But this strategy is suboptimal for smaller $C_{SR}$. The benefits of state transmission to the destination are thus clear from this example. Next, we consider state cooperation only, that is, $C_{SR} = 0$, and compare the achievable rate by our scheme in Fig. 2 (b) with the upper bound (31). We also plot the achievable rate $C^G_{\text{no coop}}$ in (30) that is attained when the source transmits message only. The benefits of cooperative state transmission by the source are clear from the figure. Moreover, if $C_{RS}$ is large enough, the scheme proposed is seen to achieve the upper bound, as proved in Corollary 3.

We get further insights into system performance by letting $N_0 \ne 0$. For message cooperation only, i.e., $C_{RS} = 0$, Fig. 3 (a) shows the rates achievable by our scheme and by the same scheme when the relay ignores the state information (29) versus signal-to-noise ratio $\gamma$. It can be seen that in general state transmission from the relay can provide rate improvement, as also shown in Fig. 2 (a). With $C_{SR}$ increasing, the achievable rate increases until it saturates at the upper bound (12) when $C_{SR}$ is large enough, e.g., when $C_{SR} = 1.2$, the achievable rate overlaps with the upper bound. For state cooperation only, that is, $C_{SR} = 0$, Fig. 3 (b) shows the rate achievable by our scheme. The upper bound therein also refers to (12). It can be seen that cooperative state transmission by the source is general advantageous, as compared to the performance without cooperation, i.e., $C_{RS} = 0$. However, unlike the case of message cooperation only, even if $C_{RS}$ is large enough, e.g., $C_{RS} = 100$ in Fig. 3 (b), the upper bound is not achievable in general. This is unlike the noiseless case shown in Fig. 2 (b), due to the fact that noise makes the state information at the destination less valuable.

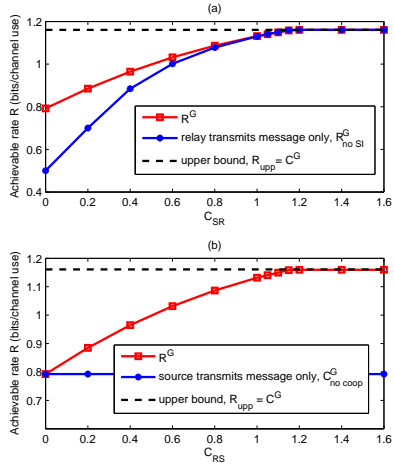
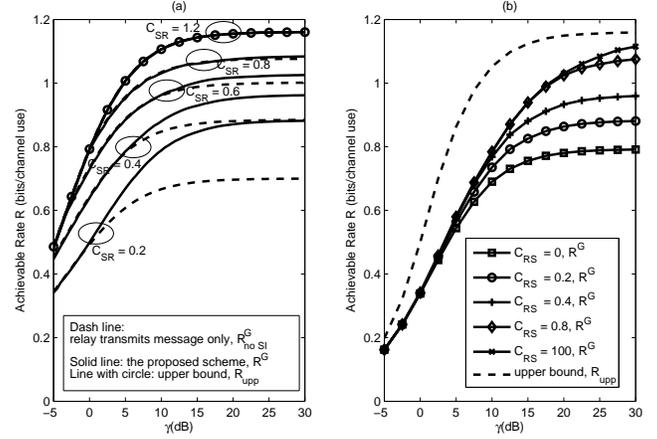

Fig. 2. Comparison of achievable rates for the case $N_0 = 0$ with $P = P_R = P_S = 1$. Fig. 2 (a) plots the achievable rates versus $C_{SR}$ for message cooperation only ($C_{RS} = 0$), and Fig. 2 (b) plots the achievable rates versus $C_{RS}$ for state cooperation only ($C_{SR} = 0$).

Fig. 3. Comparison of achievable rates for the case $N_0 \neq 0$ with $P = P_R = P_S = 1$, $\gamma = 10 \log_{10} (1/N_0) \, (dB)$. Fig. 3 (a) plots the achievable rates versus $\gamma$ for message cooperation only, in which $C_{SR} = \{0.2, 0.4, 0.6, 0.8, 1.2\}$, $C_{RS} = 0$. Fig. 3 (b) plots the achievable rates versus $\gamma$ for state cooperation only, in which $C_{SR} = 0$, $C_{RS} = \{0, 0.2, 0.4, 0.8, 100\}$.

## VI. CONCLUDING REMARKS

In this work, we have focused a state-dependent relay channel where state information is available at the relay in a strictly causal fashion. Assuming that source and relay can communicate via conferencing links, cooperation is enabled for both transmission of message and state information to the destination. First, we have proposed an achievable coding scheme that exploits both message and state cooperation. Next, capacity results have been established for some special cases. Finally, we have briefly considered the Gaussian model and obtained some capacity results. In general, our results point to the advantage of state information at the relay, despite it being known only strictly causally. This is unlike point-to-point channels. Moreover, for given conferencing capacities, both state and message cooperation can in general improve the achievable rate.